\documentclass{PoS}

\usepackage{amsmath}
\usepackage{bm}

\newcommand{\op}{\ensuremath{\mathcal{O}}}
\newcommand{\latop}{\ensuremath{O}}
\newcommand{\renop}{\ensuremath{\bar{O}}}
\newcommand{\MSbar}{\ensuremath{\overline{\text{MS}}}}
\newcommand{\mx}[1]{\ensuremath{\langle #1 \rangle}}
\newcommand{\CKM}{CKM}
\title{$\bm{D}$-Meson Mixing in {2+1}-Flavor Lattice QCD} 

\ShortTitle{$D$-Meson Mixing}

\author{Chia~Cheng~Chang$^{abc}$\thanks{Present address: Lawrence Berkeley National Laboratory, 
        Berkeley, California, 94720, USA}\;,
    C.~M.~Bouchard$^{c}$\thanks{Present address: School of Physics and Astronomy,
        University of Glasgow, Glasgow G12~8QQ, UK}\;,
    A.~X.~El-Khadra$^{ab}$,
    E.~Freeland$^d$,
    E.~G\'amiz$^e$,
    \speaker{A.~S.~Kronfeld}$^{\;bf}$, 
    J.~W.~Laiho$^g$,
    E.~T.~Neil$^{hi}$,
    J.~N.~Simone$^b$, and 
    R.~S.~Van~de~Water$^b$ \\
    \llap{$^a$}Department of Physics, University of Illinois, Urbana, Illinois, 61801, USA \\
    \llap{$^b$}Fermi National Accelerator Laboratory, Batavia, Illinois, 60510, USA \\
    \llap{$^c$}Physics Department, The College of William and Mary, Williamsburg, Virginia, 23185, USA \\
    \llap{$^d$}Liberal Arts Department, School of the Art Institute of Chicago, Chicago, Illinois, 60603, USA \\
    \llap{$^e$}CAFPE and Departamento de F\'{\i}sica Te\'orica y del Cosmos, Universidad de Granada, 
    18071~Granada, Spain \\
    \llap{$^f$}Institute for Advanced Study, Technische Universit\"at M\"unchen, 85748~Garching, Germany \\
    \llap{$^g$}Department of Physics, Syracuse University, Syracuse, New York 13244, USA \\
    \llap{$^h$}Department of Physics, University of Colorado, Boulder, Colorado 80309, USA \\
    \llap{$^i$}RIKEN-BNL Research Center, Brookhaven National Laboratory, Upton, New York 11973, USA \\
    E-mail: \email{chiachang@lbl.gov,ask@fnal.gov}}
\author{Fermilab Lattice and MILC Collaborations}

\abstract{
    We present results for neutral $D$-meson mixing in 2+1-flavor lattice QCD.
    We compute the matrix elements for all five operators that contribute to $D$ mixing at short distances,
    including those that only arise beyond the Standard Model.
    Our results have an uncertainty similar to those of the ETM collaboration (with 2 and with 2+1+1
    flavors).
    This work shares many features with a recent publication on $B$ mixing and with ongoing work on 
    heavy-light decay constants from the Fermilab Lattice and MILC Collaborations.}

\FullConference{34th annual International Symposium on Lattice Field Theory\\
		24-30 July 2016\\
		University of Southampton, UK}

\begin{document}

\section{Introduction}
\label{sec:intro}

These proceedings contain a status update of an ongoing calculation of $D^0$-$\bar{D}^0$ mixing matrix
elements~\cite{Chang:2014cea}, similar to our published work on $B^0$-$\bar{B}^0$
mixing~\cite{Bazavov:2016nty}.
We present nearly final results for all five matrix elements, sufficient to describe $D^0$-$\bar{D}^0$
mixing not only in the Standard Model, but also in any high-energy extension that modifies only the local
$\Delta C=2$ interaction.

\begin{figure}[b]
    \centering
    \vspace*{-1em}
    \includegraphics[width=0.7\textwidth]{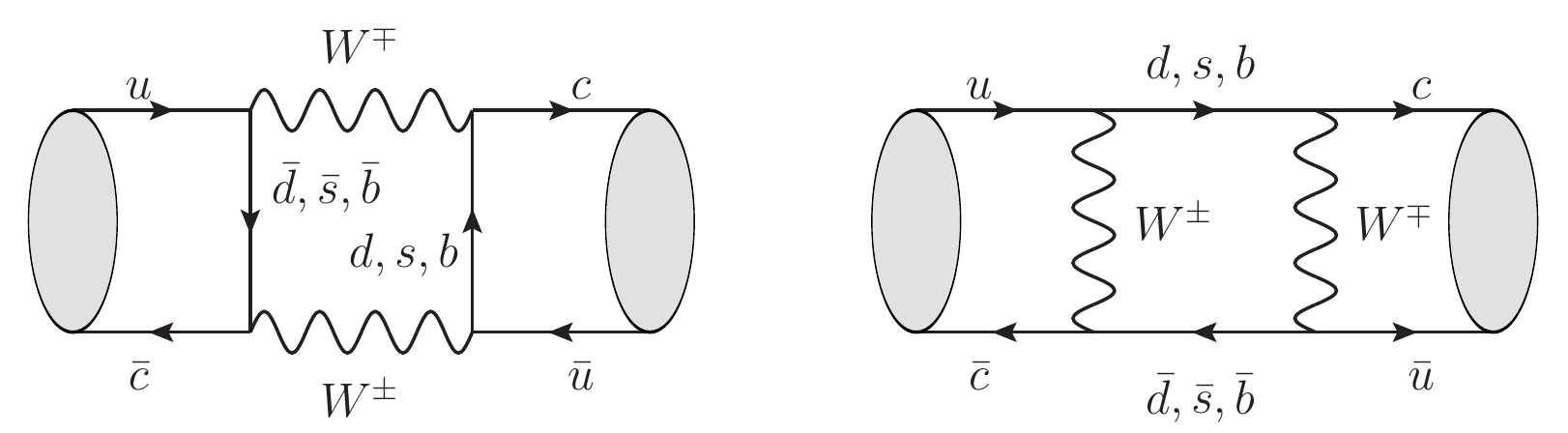}
    \vspace*{-1em}
    \caption{Box diagrams mediating  $D^0$-$\bar{D}^0$ mixing in the Standard Model.}
    \label{fig:box}
\end{figure}

In the Standard Model, neutral-meson mixing is mediated by one-loop, GIM-suppressed processes, shown in
Fig.~\ref{fig:box}.
In extensions of the Standard Model, other particles could appear in the boxes; there could even be
tree-level flavor-changing neutral currents.
Mixing has been observed in all four neutral-meson systems---$K^0$, $D^0$, $B^0$, and $B^0_s$---but the
pattern of internal quark masses and \CKM\ factors explains why the phenomenology
differs so greatly from one system to another.

Because the $W$ bosons and $b$ quarks have masses well above the QCD scale, mixing can be
re-expressed as stemming both from a local $\Delta C=2$ interaction and two $\Delta C=1$ interactions
separated by a distance of order $1/\Lambda_\text{QCD}$.
From degenerate perturbation theory, the off-diagonal term in the mass-width matrix is~\cite{Artuso:2008vf}
\begin{equation}
    M_{12} - \frac{i}{2}\Gamma_{12} \propto \langle D^0|\mathcal{L}^{\Delta C=2}|\bar{D}^0\rangle +
        \sum_n \frac{\langle D^0|\mathcal{L}^{\Delta C=1}|n\rangle\langle n|\mathcal{L}^{\Delta C=1}|\bar{D}^0\rangle}{M_D-E_n+i0^+}.
    \label{eq:mix}
\end{equation}
The second term is very difficult to estimate.
For $D^0$ mesons it is also not negligible, unlike for $B^0$ and $B^0_s$, where $t$, $c$, and~$u$ quarks
appear in the box.
(For kaons, the second term is important but not dominant.) \@ %
One can relate the measured mass and width differences, $\Delta M$ and $\Delta\Gamma$, to $|M_{12}|$,
$|\Gamma_{12}|$, and the relative phase $\arg(\Gamma_{12}/M_{12})$~\cite{Buras:1984pq}.
In some extensions of the Standard Model, only the first term and, thus, $M_{12}$
is altered~\cite{Golowich:2009ii}.

The effective Lagrangian $\mathcal{L}^{\Delta C=2}$ (at energies below the $b$-quark mass) is built out of
the following operators (and their Wilson coefficients)~\cite{Hagelin:1992tc,Gabbiani:1996hi,Bagger:1997gg}:
\begin{align}
    \op_1 &= \bar{c}\gamma^\mu Lu\,\bar{c}\gamma_\mu Lu, &
        \tilde{\op}_1 &= \bar{c}\gamma^\mu Ru\,\bar{c}\gamma_\mu Ru,
    \label{eq:O1} \\
    \op_2 &= \bar{c} Lu\,\bar{c} Lu, &
        \tilde{\op}_2 &= \bar{c} Ru\,\bar{c} Ru,
    \label{eq:O2}  \\
    \op_3 &= \bar{c}^\alpha Lu^\beta\,\bar{c}^\beta Lu^\alpha, &
        \tilde{\op}_3 &= \bar{c}^\alpha Ru^\beta\,\bar{c}^\beta Ru^\alpha,
    \label{eq:O3}  \\
    \op_4 &= \bar{c} Lu\,\bar{c} Ru,
    \label{eq:O4}  \\
    \op_5 &= \bar{c}^\alpha Lu^\beta\,\bar{c}^\beta Ru^\alpha,
    \label{eq:O5}
\end{align}
where $L$ ($R$) denotes a left-(right-)handed projector on the Dirac indices, and $\alpha$ and $\beta$ are
color indices.
By parity conservation in QCD, %
$\langle D^0|\tilde{\op}_i|\bar{D}^0\rangle=\langle D^0|\op_i|\bar{D}^0\rangle$, $i=1$, 2,~3.
Thus, the five matrix elements $\langle D^0|\op_i|\bar{D}^0\rangle$, $i=1,\ldots5$, suffice to
describe the short-distance part of all $\Delta C=2$ processes, whether their origin is $W$-$b$ box or
something else.
In these proceedings, we report on a calculation of all five matrix elements using lattice QCD with 2+1
flavors of sea quarks.


\section{Lattice-QCD calculation}
\label{sec:qcd}

Our $D$-meson calculations have much in common with our published $B$-meson work~\cite{Bazavov:2016nty}.
We use the same ensembles (generated by the MILC collaboration) with 2+1 flavors of sea
quark~\cite{Bazavov:2009bb}.
The light quarks (valence and sea) are based on the staggered asqtad action; the heavy $c$ (or $b$) quark on
the Fermilab interpretation of the clover action.
The lattice spacings for the ensembles satisfy $a\approx0.045$~fm, $\approx0.06$~fm, $\approx0.09$~fm, and
$\approx0.12$~fm.
The sea-quark masses yield pions with
\begin{eqnarray}
    177~\text{MeV} \lesssim & M_\pi            & \lesssim 555~\text{MeV}, \label{eq:MpiG} \\
    257~\text{MeV} \lesssim & M_\pi^\text{rms} & \lesssim 670~\text{MeV}, \label{eq:Mpi-rms}  
\end{eqnarray}
The ensembles contain 600--2200 gauge-field configurations, and we use 4 or 8 sources/config.

To carry out the chiral-continuum extrapolation, we take into account the subtle way in which spin emerges
for staggered fermions with staggered-Wilson four-fermion lattice operators.
The three-point correlation function, it turns out, contains contributions not only from the continuum-limit
operator of desired spin, but also some of the wrong spin~\cite{Bernard:2013dfa}.
Because only the five operators in Eqs.~(\ref{eq:O1})--(\ref{eq:O5}) can arise, we automatically have the
information needed to disentangle this effect.
We use the one-loop chiral-perturbation-theory formulas of Ref.~\cite{Bernard:2013dfa} to remove the
wrong-spin contribution in the course of our chiral-continuum fit.

The operators in Eqs.~(\ref{eq:O1})--(\ref{eq:O5}) require renormalization for any ultraviolet regulator.
We carry out the renormalization of the lattice operators corresponding to Eqs.~(\ref{eq:O1})--(\ref{eq:O5})
together with matching to \MSbar\ schemes in continuum~QCD.
We use a mostly nonperturbative method to handle the largest lattice-to-continuum matching
corrections~\cite{ElKhadra:2001rv,Harada:2001fi}, supplemented with a one-loop calculation of the remaining,
small renormalization parts~\cite{Evans:2009du,Bazavov:2016nty}.
We choose the renormalization scale for $D$-meson matrix elements to be $3~\text{GeV}$, while we chose $m_b$
for $B_{(s)}$ mesons.

\begin{figure}[b]
    \centering
    \vspace*{-0.125em}
    \includegraphics[width=0.8\textwidth,trim=0.75in 2.65in 0.75in 2.65in,clip]{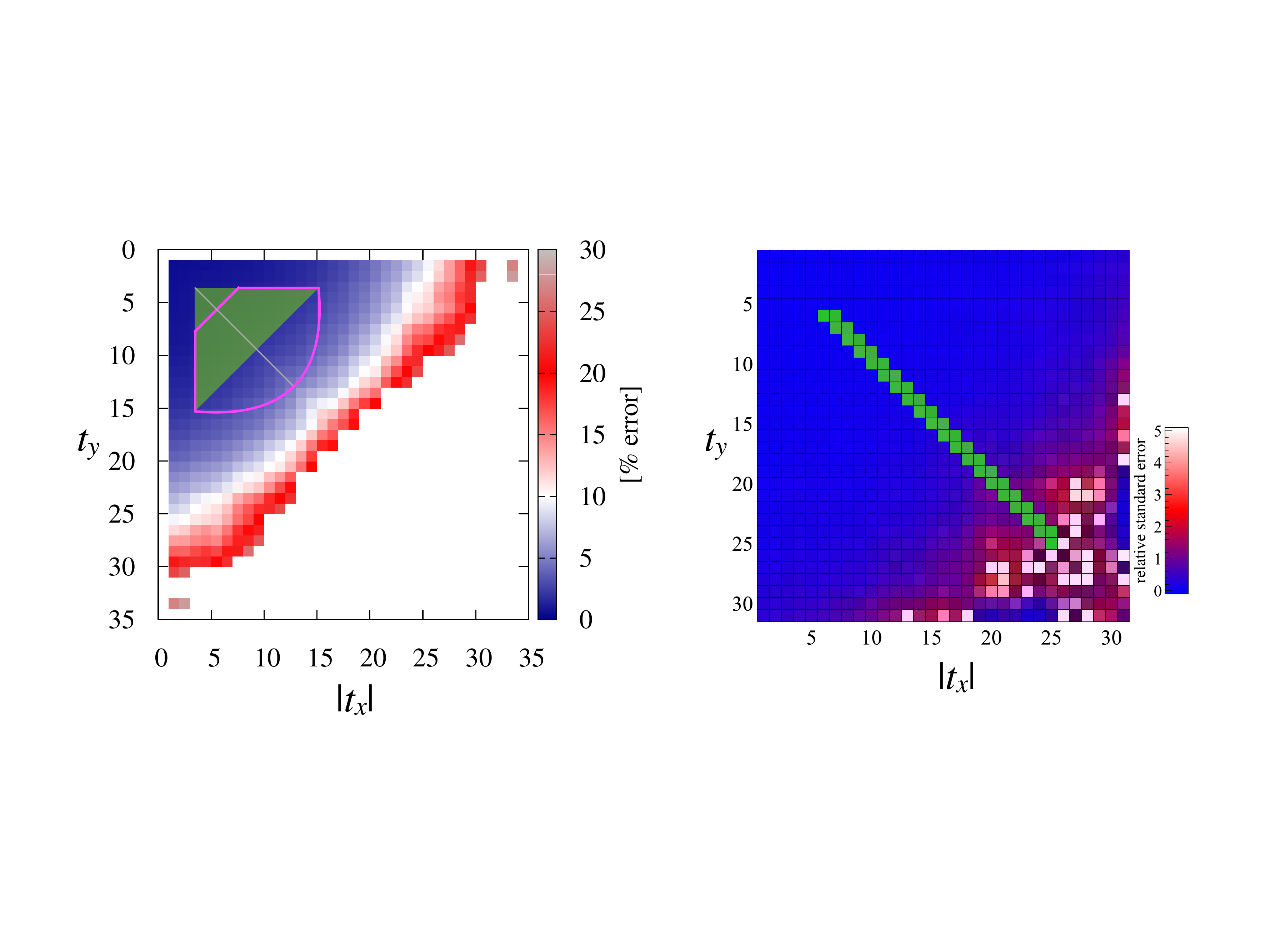}
    \vspace*{-0.125em}
    \caption{Fitting ranges for three-point correlators: triangular (green) and-or fan-shaped (magenta)
        regions for $B$ mixing (left); two-strip diagonal region (green) for $D$ mixing (right).
        Background color shows the signal-to-noise ratio from good (blue) to bad (red).}
    \label{fig:3pt}
\end{figure}
The main difference between our work on $D$ vs.\ $B_{(s)}$ mesons is the analysis of the correlation
functions.
The signal-to-noise ratio is much better for $D$-meson correlators.
For the two-point correlators, the optimal time range $t_\text{min}\lesssim t\lesssim t_\text{max}$ differs:
$t_\text{min}\approx0.7\,(0.2)$~fm, $t_\text{max}\approx3.0\,(2.4)$~fm for $D$ ($B_{(s)}$) mesons.
The difference for the three-point correlators is more striking.
We fix the four-quark operators at $t=0$ and the meson creation (annihilation) operator at time $t_x<0$
\pagebreak ($t_y>0$).
As shown in Fig.~\ref{fig:3pt}, we use a triangular and-or fan-shaped region in the $|t_x|$-$t_y$ plane for
$B_{(s)}$ mesons~\cite{Bazavov:2016nty}, while we use a long diagonal of width~2 for $D$ mixing,
$\{|t_x|=t_y\}\cup\{|t_x|=t_y+1\}$.
The long diagonal makes it easier to disentangle the lowest-lying state, if the signal persists that far.
A simultaneous fit to two- and three-point functions is used to extract the matrix elements
$\mx{\latop_i}\equiv\langle D^0|\latop_i|\bar{D}^0\rangle$.

\section{Chiral-continuum extrapolation}
\label{sec:ch-cont}

\begin{figure}[b]
    \centering
    \includegraphics[height=0.475\textheight]{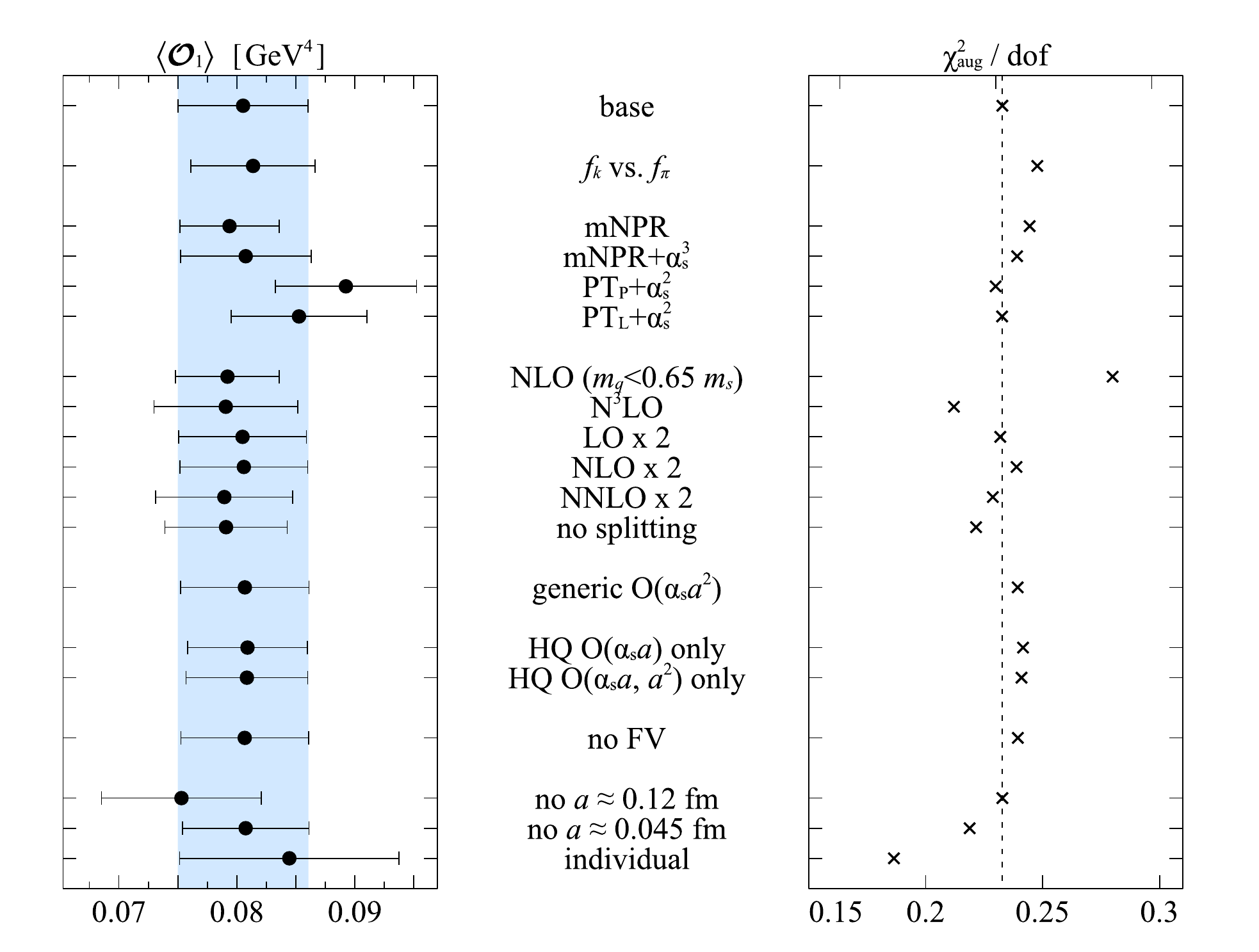}
    \caption[fig:stab]{Stability of the chiral-continuum extrapolation for several variants of the fit 
        function~$F_i$: \linebreak
        $\mx{\op_1}$~(left), minimized $\chi^2_\text{aug}/\text{dof}$ (right).
        Stability plots for the other $\mx{\op_i}$ look similar.}
    \label{fig:stab}
\end{figure}

To carry out the chiral-continuum extrapolation, we develop a fit function based on chiral perturbation 
theory ($\chi$PT), Symanzik effective field theory, and heavy-quark effective theory (HQET).
It takes the form
\begin{equation}
    F_i = F_i^\text{logs} + F_i^\text{analytic} + F_i^\text{HQ disc} + F_i^{\alpha_sa^2~\text{gen}} +
        F_i^\text{renorm} + F_i^\kappa ,
    \label{eq:Fi}
\end{equation}
where $F_i^\text{logs}$ denotes the next-to-leading order description from heavy-meson rooted staggered
$\chi$PT, with nonanalytic terms including those that disentangle the wrong-spin 
contributions~\cite{Bernard:2013dfa}; %
$F_i^\text{analytic}$ is a polynomial of various terms that arise in $\chi$PT at next-to-leading or higher 
order;
$F_i^\text{HQ disc}$ describes heavy-quark discretization effects using HQET as a theory of cutoff 
effects~\cite{Harada:2001fi}; %
$F_i^{\alpha_sa^2~\text{gen}}$ parametrizes generic cutoff effects of light quarks and gluons,
\`a~la~Symanzik; %
and $F_i^\text{renorm}$ allows the fit to be sensitive to higher orders in $\alpha_s$ for matching and 
renormalization.
Finally, $F_i^\kappa$ incorporates a correction for tuning the charm-quark hopping parameter~$\kappa$, 
based on extra runs at $a\approx0.12$~fm.

Both the renormalization and wrong-spin effects mix operators 1, 2, and~3 with each other, and also 4 and~5
with each other.
It is thus natural to fit the matrix elements in each sector simultaneously.
Some ingredients in $F_i^\text{logs}$ are common for all~$i$, such as masses, $f_\pi$, light-meson $\chi$PT
constants~\cite{Aubin:2004fs}, and the $D^*$-$D$-$\pi$ coupling.
We introduce these external inputs with Gaussian priors, for example $g_{D^*D\pi}=0.53\pm0.8$.
Because of these common ingredients, we choose to fit all five matrix elements simultaneously.
We form a $\chi^2$ function from $F_i-\mx{\renop_i}$ and the sample covariance matrix of the
$\mx{\renop_i}$, where $\renop_i$ denotes the renormalized lattice operators (which differ from the 
continuum $\op_i$ by discretization effects and higher-order matching effects).
We then augment this $\chi^2$ with Gaussian priors for the fit parameters implied in Eq.~(\ref{eq:Fi}),
choosing a central value of 0 and width of $\pm1$ in natural units for $\chi$PT and
HQET~\cite{Bazavov:2011aa} and minimize the resulting~$\chi^2_\text{aug}$.
We reconstitute the fit function at zero lattice spacing and physical quark masses to obtain our estimate of 
the $\mx{\op_i}$ and their uncertainty.

We have 510 data points for $\mx{\renop_i}$, ranging over the ensembles, valence-quark masses, and five
operators.
In our base version of $F_i$, there are 127 parameters.
To check whether the final results are robust, we repeat the procedure with several variants of~$F_i$, as
illustrated in Fig.~\ref{fig:stab}.
We express the $\chi$PT with $f_K$ instead of $f_\pi$; we choose different orders of $\alpha_s$ in
$F_i^\text{renorm}$ and even replace the mostly nonperturbative (mNPR) matching with a fully perturbative
(PT) one; we check various alternatives for the polynomial~$F_i^\text{analytic}$ (NLO, NNLO, N$^3$LO); we
check what happens when the $\chi$PT prior widths in~$F_i^\text{analytic}$ are doubled; we check
alternatives for the heavy-quark discretization errors; we substitute infinite-volume one-loop integrals for
the finite-volume sums in one-loop $\chi$PT; we omit the data from the coarsest or finest lattice spacing;
we fit each matrix element separately, thereby ignoring data constraints on wrong-spin contributions.
As one can see from Fig.~\ref{fig:stab}, the results for the $\mx{\op_1}$ are very stable, so we take these
variations in the fit as cross checks.
The same applies to the other $\mx{\op_i}$.
The largest deviations are $\sim1\sigma$ and come from fits that omit important information.
Our nearly final results for $D$ mixing are given in Table~\ref{tab:results}, together with published
results for $B_{(s)}$ mixing from Ref.~\cite{Bazavov:2016nty}.
\begin{table}[t]
    \centering
    \newcommand{\h}{\hphantom{-}}
    \newcommand{\p}{\hspace*{1.25em}\hphantom{-}}
    \newcommand{\m}{\hspace*{1.25em}-}
    \begin{tabular}{clll}
        \hline\hline
    BBGLN~\cite{Beneke:1998sy} & \multicolumn{1}{r}{$\mx{\op_i}/M_D~\left(\text{GeV}^{3}\right)$} &
    \multicolumn{2}{c}{$f_{B_q}^2B_{B_q}~\left(\text{GeV}^2\right)$} \\
                    &                           & \multicolumn{1}{c}{$q=d$} & \multicolumn{1}{c}{$q=s$} \\
        \hline
    $\mathcal{O}_1$ &     $\p0.0432(29)(9)$     &       $0.0342(29)(7)$     &       $0.0498(30)(10)$    \\
    $\mathcal{O}_2$ &     $\m0.0833(38)(17)$    &       $0.0303(27)(6)$     &       $0.0449(29)(9)$     \\
    $\mathcal{O}_3$ &     $\p0.0248(16)(5)$     &       $0.0399(77)(8)$     &       $0.0571(77)(11)$    \\
    $\mathcal{O}_4$ &     $\p0.1469(69)(30)$    &       $0.0390(28)(8)$     &       $0.0534(30)(11)$    \\
    $\mathcal{O}_5$ &     $\p0.0554(38)(11)$    &       $0.0361(35)(7)$     &       $0.0493(36)(10)$    \\
    $\mu$           & \multicolumn{1}{c}{3~GeV} & \multicolumn{1}{c}{$m_b$} & \multicolumn{1}{c}{$m_b$} \\   
        \hline\hline
    \end{tabular}
    \caption[tab:results]{Results for $D$ [this work] and $B$~\cite{Bazavov:2016nty} mixing in the 
        renormalization scheme of Ref.~\cite{Beneke:1998sy}.}
    \label{tab:results}
\end{table}
These matrix elements (as noted above) depend on the renormalization scheme; the tabulated results are in 
the \MSbar\ scheme with naive (fully commuting)~$\gamma^5$ and the evanescent-operator basis used by Beneke, 
Buchalla, Greub, Lenz, and Nierste (BBGLN)~\cite{Beneke:1998sy}.


The MILC asqtad ensembles omit the charmed-quark sea.
As in Ref.~\cite{Bazavov:2016nty}, we assign an additional 2\% uncertainty to account for this omission.
This uncertainty is given separately, in the second set of parentheses, in Table~\ref{tab:results}.

\section{Outlook}
\label{sec:outro}

Our results agree well with and have similar uncertainty as previous lattice-QCD results from the ETM
collaboration, with 2~\cite{Carrasco:2014uya} or 2+1+1~\cite{Carrasco:2015pra} flavors in the sea.
The comparison of these results tests not only the flavor-dependence of the matrix elements but also the
sensitivity to lattice fermion formulation: ETM employs twisted-mass Wilson fermions, while we employ
staggered fermions.
All these calculations use several lattice spacings and take the continuum limit.
References~\cite{Carrasco:2014uya,Carrasco:2015pra} report the so-called ``bag factors'' often used in
phenomenology~\cite{Gabbiani:1996hi}; a detailed comparison would require choices of quark masses and decay
constants (and their uncertainties) that would obscure the error budget of one or the other set of results.
We have a set of calculations underway~\cite{Kronfeld:2015xka} to compute the $D$- and $B_{(s)}$-meson decay
constants on the same ensembles and will report the bag factors then.

Estimates of the contribution to $M_{12}$ of the second term in Eq.~(\ref{eq:mix}) range over
$(10^{-3}\text{--}10^{-2})\Gamma$ \cite{Falk:2004wg}, where $\Gamma$ is the total width of the neutral $D$
meson.
It turns out, however, that all Standard-Model phases appearing in Eq.~(\ref{eq:mix}) are small.
Thus, in a TeV-scale model that might produce a large phase in $M_{12}$, the results for the $\mx{\op_i}$
can be used to constrain the model's parameters.
Furthermore, until a method is developed to tame the second term in Eq.~(\ref{eq:mix}), the accuracy
achieved in this work and Refs.~\cite{Carrasco:2014uya,Carrasco:2015pra} should suffice for this purpose.

\acknowledgments

We thank our collaborators in the Fermilab Lattice and MILC Collaborations.
Computations for this work were carried out with resources provided by the USQCD Collaboration, the National
Energy Research Scientific Computing Center, and the Argonne Leadership Computing Facility, which are funded
by the Office of Science of the U.S.\ Department of Energy; and with resources provided by the National
Institute for Computational Science and the Texas Advanced Computing Center, which are funded through the
National Science Foundation's Teragrid/XSEDE Program.
This work was supported in part by the U.S.\ Department of Energy under grants
No.~DE-FG02-13ER42001 (C.C.C., A.X.K.), No.~DE-SC0015655 (A.X.K.),
No.~DE-SC0010005 (E.T.N.);
by the U.S.\ National Science Foundation under grant 
PHY14-17805~(J.L.); 
by the Fermilab Fellowship in Theoretical Physics (C.M.B., C.C.C.);
by the URA Visiting Scholars' program (C.M.B., C.C.C., A.X.K.);
by the MICINN (Spain) under grant FPA2010-16696 (E.G.); by the Junta de Andaluc\'ia (Spain) under Grants 
No.~FQM-101 and No.~FQM-6552 (E.G.);
by the European Commission (EC) under Grant No.~PCIG10-GA-2011-303781 (E.G.);
by the German Excellence Initiative and the European Union Seventh Framework Program under grant agreement 
No.~291763 as well as the European Union's Marie Curie COFUND program (A.S.K.).
Fermilab is operated by Fermi Research Alliance, LLC, under Contract No.\ DE-AC02-07CH11359 with the United
States Department of Energy.
Brookhaven National Laboratory is supported by the United States Department of Energy under contract No.\
DE-SC0012704.

\end{document}